# Particle Effects on the ISGRI Instrument On-Board the INTEGRAL Satellite

A. Claret, O. Limousin, F. Lugiez, P. Laurent and M. Renaud

*Abstract*—The *INTEGRAL* satellite was launched on October 17, 2002. All on-board instruments are operating successfully. In this paper, we focus on radiation effects on the Cadmium Telluride camera ISGRI. The spectral response of the camera is affected by cosmic particles depositing huge amount of energy, greater than the high threshold of the electronics. Our study raises the contribution of cosmic ray protons. Solutions are proposed to limit the degradation of spectral response of large pixel gamma cameras operating in space.

## I. INTRODUCTION

THE INTEGRAL (INTErnational Gamma-Ray Astrophysics Laboratory) satellite is an ESA (European Space Agency) gamma-ray observatory successfully launched from Baïkonour on October 17, 2002. It is devoted to the observation of the universe between few keV up to 10 MeV, the energy range of the most extreme phenomena (black hole, accretion in X-ray binaries, gamma-ray bursts, supernovae...). Two telescopes are mounted on INTEGRAL. The first one, IBIS (Imager on Board the Integral Satellite), provides diagnostic capabilities of fine imaging (12 arcmin full width half maximum) whereas the second one, SPI (SPectrometer on Integral), performs spectral analysis of gamma-ray sources. Both are coded mask aperture telescopes. In order to achieve a spectral coverage from several tens of keV to several MeV, the detection unit of the imager IBIS [1] is composed by two gamma cameras, ISGRI (Integral Soft Gamma-Ray Imager) [2] covering the range from 15 keV to 1 MeV and PICsIT (Pixilated CsI Telescope) [3] covering the range from 170 keV to 10 MeV. This paper deals with the ISGRI cadmium telluride (CdTe) camera, the low energy position sensitive detector of IBIS.

The in-flight behavior of ISGRI was already reported by Limousin et al. [4]. After a short description of the camera, its energy response is presented in Section II. Effects of radiations on ISGRI are addressed in Section III. But one has to keep in mind that there are two separate problems: (a) the production of activation lines after the perigee passage due to the high radiation fluence mainly coming from trapped particles in the van Allen belts, and (b) the response variation due to the impact of single particle from galactic cosmic rays or solar particles along the scientific part of the orbit, i.e. outside the van Allen belts. The first problem (a) has an influence on the spectral calibration and sensitivity of the imager. The second problem (b) has an influence on the detector stability but also on its spectral response. In this paper, we address particle effects on the spectral response of ISGRI (Sections IV and V). Some solutions for limiting the degradation of the spectral response are proposed in the conclusion (Section VI).

## II. ISGRI CAMERA

### A. General description

Energy deposits from cosmic particles in detectors operating in space affect their performances. This is particularly true for large monolithic detector (such as Anger camera) for which the spatial resolution is strongly degraded by particle impacts. Pixel gamma cameras, where each pixel is an independent detector with its own electronic chain, avoid this problem since the average time between two successive impacts in a single detector can be relatively long, allowing a complete recovery of the electronics. This led to the choice of using CdTe crystals to build the large gamma camera ISGRI. 8 independent modular detection units compose the ISGRI detection plane. Each pixel of the camera is a CdTe detector read out by a dedicated integrated electronic channel. Altogether, there are 16,384 detectors (128x128) and as many electronic channels (Fig. 1). Each detector is a 2 mm thick CdTe:Cl THM crystal of 4 mm x 4 mm with platinum electrodes. (see [2], [4] for more details).

### B. Energy response

To determine accurately the energy of incoming photons in ISGRI, a simultaneous measurement of the standard pulse-height and the pulse rise-time for every event is necessary. The amplitude of the pulse depends on the energy of interacting photons, whereas the rise-time of the pulse depends on the interaction depth of photons in the detector because of the charge carrier properties in the CdTe. Let's consider a fixed energy of the incident photon: if it is stopped near the surface of the crystal (cathode side), then the signal mainly due to the electron motion is fast, leading to a short rise-time of the pulse. On the contrary, if the photon is stopped deeper in the crystal the hole contribution to the signal make it slower, leading to long rise-time of pulse.

When the pulse rise-time is long, charge loss and ballistic deficit occurs, leading to a poor measurement of the energy. In order to compute an off-line correction of the charge losses and then retrieve a good sensitivity and a good spectral response, pulse rise time and pulse height are systematically recorded in a bi-parametric diagram (2D histogram of pulse rise-time and pulse-height). Fig 2 shows an example of a diagram recorded in flight.

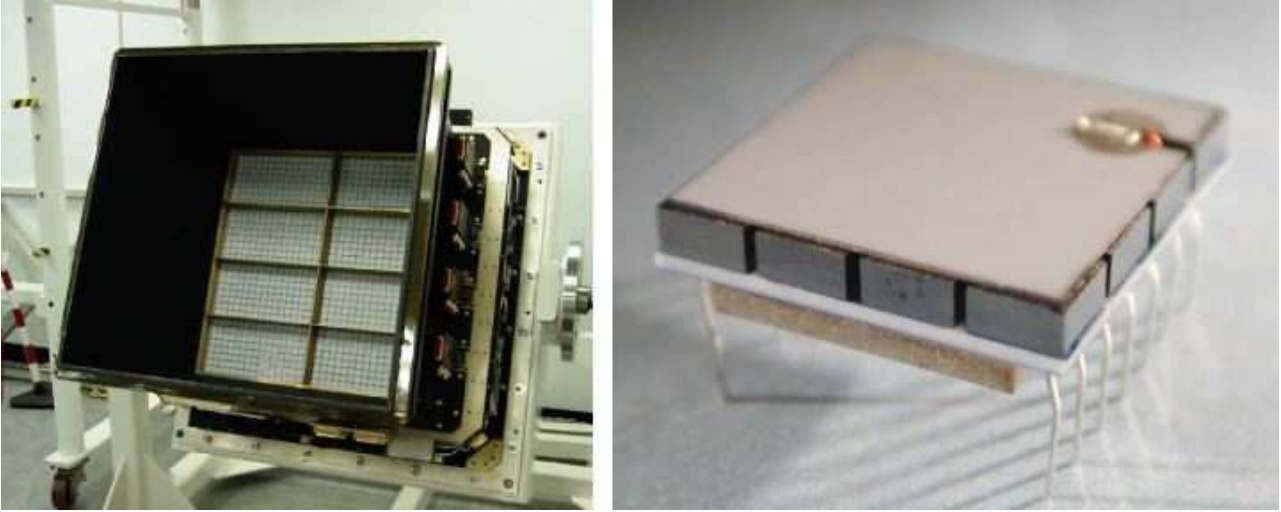

Fig. 1: Left panel is a view of the 8 ISGRI modular detection units (white color) at the bottom of the Tungsten passive shield well (black color) after integration in the IBIS detection unit. Each modular detection unit is composed of 128 polycells (8x16). Right panel represents an individual polycell made of 4x4 CdTe crystals of 2 mm thick. The readout electronics is visible at the bottom of the polycell.

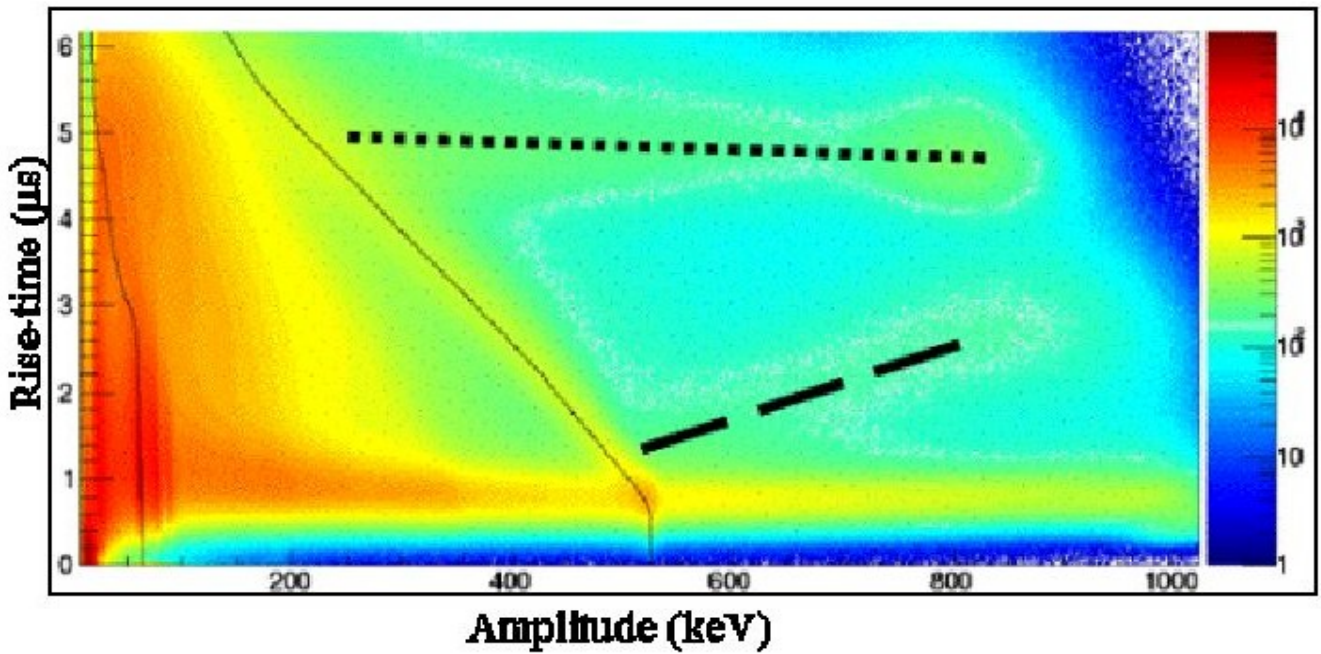

Fig. 2: In-flight bi-parametric diagram (histogram of pulse rise-time vs. pulse amplitude) obtained with the calibration unit data. The 511 keV line of the calibration source ($^{22}$Na) goes from high amplitude at low rise-time to low amplitude at high rise-time (thin solid line), depending on the penetration depth of photons into the CdTe crystals. Undesired structures (dotted and dashed lines) are also visible. They are attributed to cosmic ray particles. Note that the white contour of these undesired structures has been enhanced for illustrative purpose.

Cosmic ray particles interacting with the crystals correspond to a particular signature in the bi-parametric diagram (dotted and dashed lines in Fig. 2), different of the one described above. The energy response of ISGRI is degraded due to the presence of these undesired structures, which represent ~ 80 counts/s in bi-parametric diagrams over ~ 900 counts/s in total. This induces an increase of dead time of ~ 10%, reducing the telemetry flow of scientific data. Signal to background ratio of celestial sources is also degraded in the polluted part of bi-parametric diagram.

## III. ISGRI RADIATION ENVIRONMENT

### A. Particle environment

The orbit of *INTEGRAL* was selected to minimize the background level of instruments, maximizing at the same time its uninterrupted scientific observing time and also the telemetry flow. The chosen orbit has a period of 72 hours, a perigee of ~10,000 km, an apogee of ~155,000 km and an inclination of ~51°. The trajectory crosses the whole outer electron belt and just barely touches the inner belt with increased fluxes of protons. Outside the outer electron belt (~65 hours over 72 hours of the orbit), the mission is directly exposed to fairly constant cosmic ray background (85% protons, 12% alphas and 3% heavier nuclei) and solar particles.

### B. Particle effects

The three main effects of particle impacts on ISGRI concern:

*1) The production of activation lines*, due to the encounter of the spacecraft with the earth's radiation belts. This has an influence on the spectral calibration and sensitivity and was already addressed in [6].

*2) The detector stability*, controlled by the so-called noisy pixel handling system, which automatically switch off or raise the low threshold of noisy pixels [2]. This system was particularly useful to overcome problems due to the passage of cosmic particles in the crystals. The system had to be tuned once in flight conditions were determined (see [4] for details).

*3) The spectral response*, affected by cosmic particles depositing in the crystals huge amounts of energy (much greater than the high threshold of the electronics).

In this paper, we only focus on this last effect. The in-flight bi-parametric diagram (Fig. 2) shows an undesired high rise-time (~5 μs) structure covering almost the whole amplitude range (dotted line). This horizontal structure reduces the sensitivity for long rise-time events. Another undesired structure (dashed line) is visible in this diagram. It covers a region with pulse height increasing with rise-time between 500 keV to 1 MeV. The following study intends to determine the origin of these structures. This could help to limit these effects for future high energy instruments.

## IV. MODEL OF PARTICLE EFFECTS

### A. Basic idea

The low energy threshold of the electronics is 12 keV and the high threshold is 1 MeV. But cosmic-ray particles can induce much higher energy deposits when they interact with the CdTe crystals. Such events are not taken into account by the electronics because of the thresholds, but they can induce cross-talk effects in preamplifiers neighboring the hit pixel. What kind of particles can generate energy deposit greater than the high threshold of the electronics?

*1) Protons:* To be able to pass through the CdTe crystals with normal incidence, protons must have incident energy at least of 24 MeV. Incident energy up to 44 MeV is necessary for those passing through the diagonal of a crystal (incident angle of 70°). At higher angles, several pixels would be touched and wouldn't be recorded since each ISGRI module is able to detect only one event at a time. The energy deposited by protons is then between 1.7 MeV (for high energetic protons near the minimum ionization) and 44 MeV (for low energetic protons having the longest path in CdTe crystals). Those protons are present in cosmic rays.

*2) Ions:* The low-energy part of cosmic and solar ions is strongly attenuated because of their limited range in the IBIS passive shield, whereas the unshielded remaining part of the spectrum contributes to a negligible way since they are far less numerous than protons.

*3) Electrons:* The number of electrons strongly increases when *INTEGRAL* approaches the outer electron belt. But electrons able to produce energy deposit leading to saturation of preamplifiers (minimum 5 MeV) are very few. On the other hand, 5 MeV electrons are relativistic and generally deposit only 1.7 MeV in a 2 mm thick CdTe, not enough to produce preamplifier saturation.

Undesired structures are thus mainly attributed to rather low energy cosmic-ray protons (< 100 MeV) present all along the *INTEGRAL* orbit. In order to check this assumption, our model of ISGRI spectral response under proton flux is based on three studies: the electrical model, the model of bi-parametric spectral response and Monte-Carlo simulations. These models are detailed in the following sub-sections.

### B. Electrical model for cross-talk

The electrical model of ISGRI allowed the calculation of capacitive cross-talk between channels when one of them is heavily saturated after a high energy deposit. If the preamplifier of a given channel is highly saturated then a fraction of the signal is copied in a neighbor channel. Note that the neighbor channel where the fraction of signal can be copied must belong to the same polycell (4x4 crystals), since all polycells are isolated from each other (Fig. 1 right panel).

The saturation phenomenon was analyzed thanks to the electrical model. The capacitive cross-talk between channels is detailed on Fig. 3, where $C_C$, $C_L$, $C_F$, $C_P$ represent respectively the inter-pixel, liaison (AC coupling), feedback and parasitic capacitors. With $C_F = 70$ fF and saturation voltage $V_{SAT} \sim 2.5$ V (ASIC design), we derive that saturation charge corresponds to $1.1 \times 10^6$ electrons, i.e. 5 MeV deposit in a CdTe crystal. Energy deposits between 1 MeV and 5 MeV cannot produce a signal in the camera. For a non-saturated channel, the normal cross-talk with the neighbor channel is given by $F_C = C_C/C_L \sim 0.6\%$. But for a highly saturated channel ($E_{dep} > 5$ MeV), the measured cross-talk with the neighbor channel is given by $F_C = C_C/C_P$ and is expected to be between 5 and 12%. Results are summarized as follows:

- High energy deposits ($E_{dep} > 5$ MeV) from cosmic particles saturate the charge sensitive preamplifiers of the hit pixel, leading to a modification of the cross-talk between channels in the same polycell. A neighbor of the hit pixel is then able to trigger and copy the pulse shape at the entrance of its preamplifier with an attenuation factor X ~ 88-95%.

- The measured energy in the neighbor pixel corresponds then to (1-X) ~ 5-12% of $E_{dep}$. This value becomes compatible (i.e. between the low and high level thresholds) with the ISGRI energy range (12 keV - 1 MeV) and makes the phenomenon recordable on the bi-parametric diagram of the science data.

During ground tests, no cross-talk was observed for energy deposit up to 2.3 MeV (maximum energy of $^{90}$Sr test source). This explains why the undesired structures of bi-parametric diagram were discovered in-flight.

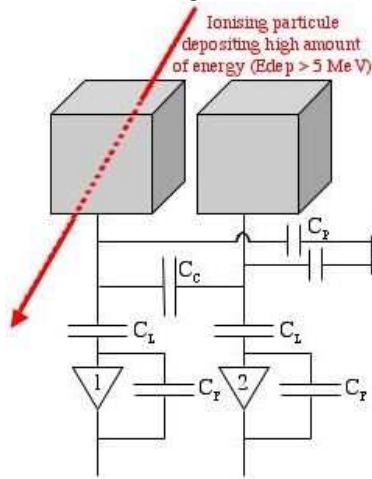

Fig. 3: Summary of the effect of high energy deposit on the neighbor pixel: first step is a saturation of preamplifier #1 ($E_{dep}$ > 5 MeV), followed by cross-talk in neighbor preamplifier #2 (5-12% of $E_{dep}$ > low energy threshold) through the inter-pixel capacity $C_C$. The other capacities $C_L$, $C_F$, $C_P$ are respectively the liaison, feedback and parasitic capacities.

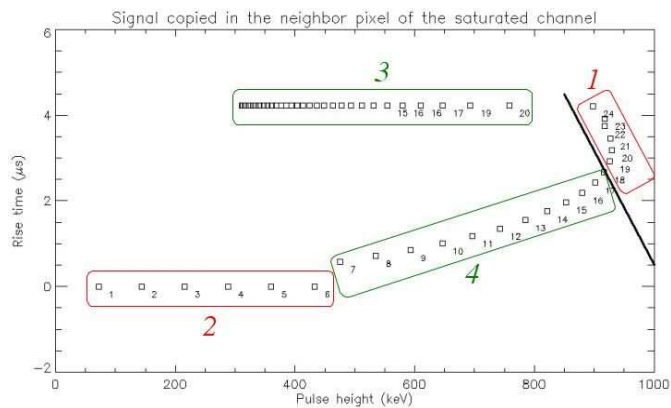

Fig. 4: Simulation of cross-talk effect in bi-parametric diagram of the neighbor pixel of the hit pixel. The numbers besides each point of the diagram correspond to the deposited energy (in MeV) by the incident proton (values correspond to normal incidence). For example, an incident proton of 17 MeV is stopped in the hit pixel, saturates the channel, then 5-12% of the deposited energy is copied in the neighbor pixel, leading to a pulse of 900 keV height and 2.4 μs rise time. The solid line corresponds to the high threshold of the electronics. See text for comparison of different regions (1, 2, 3 and 4) with structures visible on Fig. 2.

*C. Model of bi-parametric response under proton flux*

A model of the bi-parametric response of ISGRI has been elaborated to determine what is the bi-parametric response under proton flux.

The bi-parametric response model is based on the calculation of the pulse shapes out of a CdTe crystal hit by high energy protons. To evaluate the signal on the neighbor pixel, the pulse amplitudes are then attenuated by the cross-talk coefficient $F_C$ (see Section IV.B). From these pulses, pulse-height and rise-times are extracted and reported in a bi-parametric diagram.

$P(E_{dep}, t)$ (1) is the pulse shape as a function of time $t$ obtained after a particle with the initial energy $E_{inc}$ has deposited $E_{dep}(\theta, E_{inc})$ (2).

The total amount of energy $E_{dep}$ deposited by the incoming particle from the surface at $z=0$ to the depth where it stopped at $z=D$, is calculated as the sum of elementary energy deposits $dE_{dep}(z, \theta, E_{inc})$ along the path of the particle in the crystal ($z$ is the depth from the surface, $\theta$ is the incidence angle of the particle). Each elementary deposit takes into account the real energy transfer to the crystal and is tabulated.

$$P(E_{dep}, t) = \int_0^L P(dE_{dep}, z, t) \cdot dz \quad (1)$$

where $L$ is the detector thickness, $P(z)$ the elementary pulse due to local energy deposit at the depth $z$.

$$E_{dep}(\theta, E_{inc}) = \int_0^D dE_{dep}(z, \theta, E_{inc}) \cdot dz \quad (2)$$

We use the Hecht equation to determine the elementary pulse shape due to the motion of elementary charge clouds generated at every depth in the crystal by the passing particle:

$$P(dE_{dep}, z, t) = A\left[\mu_e \tau_e \left(1 - \exp\left(\frac{-t}{\tau_e}\right)\right) + \mu_h \tau_h \left(1 - \exp\left(\frac{-t}{\tau_h}\right)\right)\right] \quad (3)$$

with

$$A = \frac{qE \cdot dE_{dep}(z, \theta, E_{inc})}{L \cdot w} \quad (4)$$

where $q$ is the charge of an electron, $E$ the applied electric field to the detector, $w$ the ionization energy, $\mu_e$ and $\mu_h$ the mobility of electrons and holes, $\tau_e$ and $\tau_h$ the lifetimes of electrons and holes.

For instance, we have calculated the bi-parametric response for protons ranging from 1 to 60 MeV hitting the CdTe with a normal incidence. Results are displayed on Fig. 4. We distinguish four regions in this figure. In the region #1, the pulse copied to the pixel is so strong that, even attenuated, it goes above the high threshold value (right of the black solid line). These events occur for protons ranging from 17 to 24 MeV. They are not expected in the flight data and indeed not found. On the other hand, pulses responsible for the region #2 correspond to events not able to saturate strongly the electronics. These pulses are not expected neither. If some occur, their very low rise time (< 0.8 µs) implies they are all recorded in the first rise-time bin.

In region #4, protons can saturate the electronics and the induce signal is copied. Those protons must deposit between 7 to 17 MeV. In this range, protons always stop quite close to the surface and then produce short rise-time signals. The higher is the incident energy the deeper protons penetrate in the bulk. This explains the positive slope of region #4. The slope begins around 500 keV without any link with the famous 511 keV line. Finally, pulses of region #3 are all due to passing protons with energy higher than 24 MeV in normal incidence and all of them produce long rise-time pulses. In this region, the energy deposit is getting lower when the proton initial energy increases. Above 100 MeV, most of the protons will be vetoed and won't appear in the diagram.

In this example, the undesired structures visible on the in-flight data (Fig. 2) are clearly recognizable in region #3 and #4 of Fig 4. In the next section, we use this model through a Monte-Carlo simulation to provide a quantitative analysis.

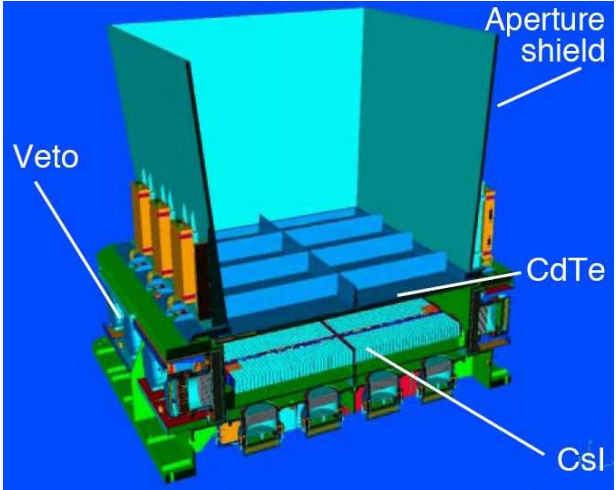

Fig 5: Cut-away view of IBIS detector assembly and passive aperture shield (hopper) as modeled in the *INTEGRAL* mass model. Layers labeled CdTe and CsI correspond respectively to ISGRI and PICsIT gamma cameras. Layer labeled Veto corresponds to anti-coincidence active shield. The coded mask at 3.2 m above plane and the passive shield are not shown.

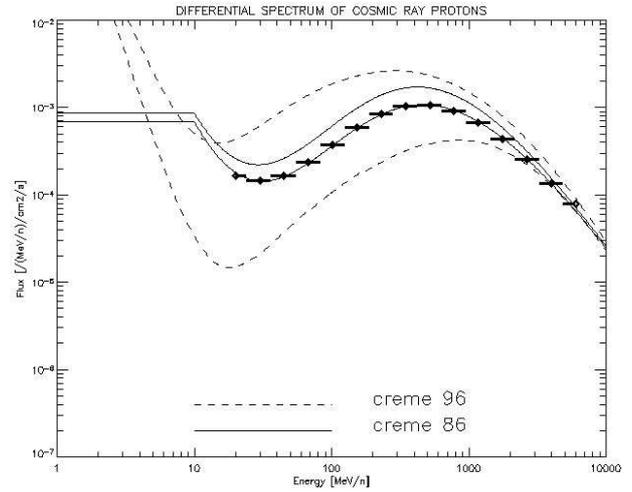

Fig. 6: Various differential spectra of cosmic ray protons used in our Monte-Carlo simulations. Spectra are represented at maximum and minimum solar activity periods: CREME 86 (solid line) and CREME 96 (dashed line). The selected energy bins between 20 MeV and 6 GeV are over plotted on the CREME 86 model at solar maximum.

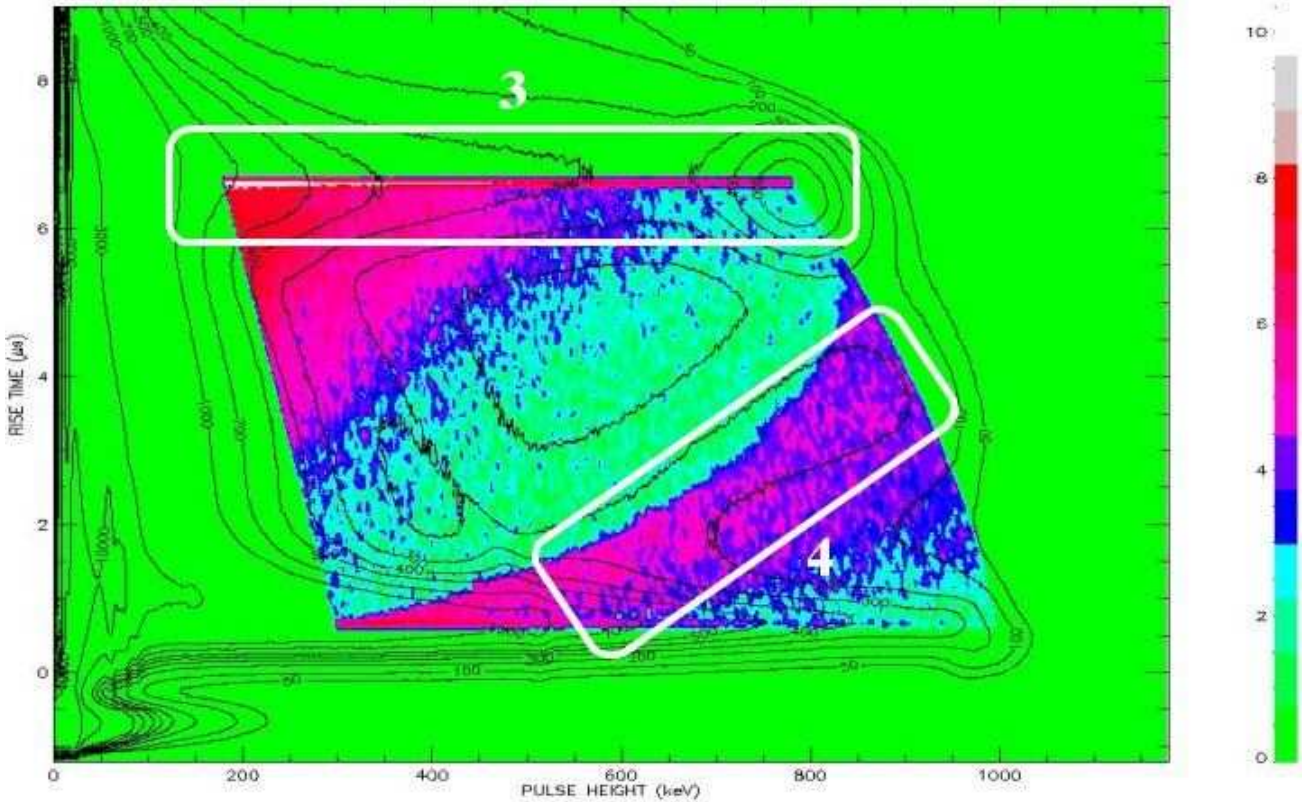

Fig. 7: Monte-Carlo simulation result for the proton spectrum derived from the CREME 86 model at solar maximum. Each color level of the bi-parametric diagram represents the contribution of an incident proton after having saturated one channel. Contour plots of the in-flight diagram are superimposed to the Monte-Carlo results for comparison. The temporal resolution of the electronics has not been modeled, which explains that the proton contribution of region #3 appears much more thin than in flight data. All numbers beside contours and color scale are in arbitrary units. The starting energy of region #4 corresponds a cross-talk factor $F_C$ of 6.2 % in well agreement with the prediction of the electrical model (5 to 12 %).

*D. Monte-Carlo simulations*

*1) General description*: Our model of ISGRI bi-parametric response under proton flux allowed us to reproduce the general shape of both undesired structures observed in flight data. It is necessary to normalize this model according to the spectrum of responsible particles (cosmic ray protons) in order to compare it quantitatively to in-flight data. To do so, it is necessary to transport the spectrum of cosmic ray protons through the various shields of the IBIS telescope. The impacts of undesired photons (those not passing through the coded mask of IBIS) are attenuated by a passive shield made of Tungsten (1 mm thick) and Lead (average thickness 1 mm). An active anti-coincidence shield (veto) made with BGO crystals (5 cm thick) also protect ISGRI from particles coming from the rear side. Cosmic particles are attenuated by this shielding before impacting CdTe crystals. Given the complex architecture of passive and active shields of IBIS, we used the full Monte-Carlo model [7] of IBIS in order to determine the actual energy deposited in ISGRI CdTe crystals by cosmic ray protons impacting the whole satellite. A cut-away view of the IBIS detection unit as modeled in our Monte-Carlo simulations is shown on Fig. 5. Main parameters of the Monte-Carlo simulations were: 15 energy bins in the 20 MeV to 6 GeV range for impacting protons; $10^7$ incident protons towards the spacecraft shielding for each energy bin; isotropic incident direction with respect to ISGRI.

*2) Cosmic ray model:* The in-flight data used in this paper were recorded during the first year of operations of *INTEGRAL*, in early 2003. We can consider that the Sun was more or less still in a period of maximum activity or that solar activity had already started to decline. So we used several spectra derived by OMERE [8] with the following options: CREME 86, M=1, solar maximum and minimum. We used also the CREME 96 spectrum [9]. Spectra are displayed on Fig. 6, together with the energy bins used in our Monte-Carlo simulations.

*3) Results:* Monte-Carlo simulation result is displayed on Fig. 7. From such simulated bi-parametric diagram, we can derive the contribution of cosmic-ray protons and compare it with the undesired structures of the in-flight diagram. Considering the uncertainties about the input spectrum of protons, we have derived the proton contribution for various models displayed on Fig. 6. Results are reported in Table 1.

## V. DISCUSSION

We conclude from Table 1 that the discrepancy between the proton contribution as deduced from Monte-Carlo simulations (Fig. 7) and the measured count rates in undesired structures in bi-parametric diagram of in-flight data (Fig. 2), is mainly due to the uncertainties on the proton flux. The ratio of predicted count rates in region #3 and #4 are quite similar to in-flight data. Thus, we think that galactic protons are indeed responsible for the undesired structures in bi-parametric diagrams. Our simulations also raised that region #3 is filled by protons of incident energy < 100 MeV, whereas region #4 is filled by protons with higher energy.

TABLE I
MONTE-CARLO RESULTS COMPARED TO IN-FLIGHT DATA
(COUNTS/S)

| Proton spectrum | Region #3 | Region #4 | TOTAL |
|---|---|---|---|
| CREME 96 solmax | 10 ± 3 | 8 ± 3 | 18 ± 4 |
| CREME 86 solmax | 22 ± 5 | 15 ± 4 | 37 ± 6 |
| CREME 86 solmin | 32 ± 6 | 22 ± 5 | 54 ± 7 |
| CREME 96 solmin | 77 ± 9 | 47 ± 7 | 124 ± 11 |
| IN-FLIGHT DATA | 50 ± 7 | 30 ± 5 | 80 ± 9 |

For future work, we should also consider a possible contribution from alpha ions. The proton flux directly measured by the IREM radiation monitor on-board *INTEGRAL* may be helpful.

## VI. CONCLUSION

We have explained the origin of undesired structures in the bi-parametric flight diagrams, which characterize the spectral response of the ISGRI camera. We raised the cosmic ray proton interactions in the CdTe crystals leading to cross-talk effects between channels. Our explanation is based on three studies: i) the electric model allowing the calculation of the excess of cross-talk between channels when one of them is heavily saturated after a high energy deposit, ii) the model of the CdTe bi-parametric response under proton flux, iii) Monte-Carlo simulations in order to normalize the two previous models according to the spectrum of cosmic ray protons.

Build gamma camera operating in space with separated pixels allows avoiding degradation of spatial resolution due to particle impacts. Nevertheless, the spectral response is still degraded by cosmic ray protons. Some solutions for limiting the degradation of the energy response of a large pixel CdTe camera operating in space such ISGRI could be:

- Measuring the pulse rise-time or penetration depth of particles in the pixel detector is helpful to remove proton contribution to the back ground.
- Recording the date and address of all triggers higher than the high threshold in order to reject events in the following few µs.
- Engineer low parasitic coupling designs.

## VII. ACKNOWLEDGMENT

The authors are very grateful to François Lebrun for fruitful discussions about the ISGRI camera. We would like also to acknowledge Régis Terrier for his contribution on the in-flight data used in this paper.